\documentclass[pra,showpacs,twocolumn,aps]{revtex4-1}

\usepackage{amsmath}
\usepackage{amssymb}
\usepackage{latexsym}
\usepackage{amsfonts}
\usepackage{epsfig}
\usepackage{psfrag}
\usepackage{graphicx}
\usepackage{color}

\newcommand{\bea}{\begin{eqnarray}}
\newcommand{\ea}{\end{eqnarray}}
\newcommand{\eea}{\end{eqnarray}}

\newcommand{\vc}[1]{\mathbf{#1}}

\begin{document}

\title{The large coordination number expansion of a lattice Bose gas at finite 
temperature}

\author{Patrick Navez$^{1}$, Friedemann Queisser$^{2,3}$
and Ralf Sch\"utzhold$^{2}$}
\affiliation{$^{1}$
Department Physics, University of Crete, PO Box 2208,
Heraklion 71003, Greece
\\
$^2$Fakult\"at f\"ur Physik, Universit\"at Duisburg-Essen, 
Lotharstrasse 1, 47057 Duisburg, Germany
\\
$^3$Department of Physics, University of British Columbia, Vancouver, 
V6T 1Z1 Canada}
%\ead{ralf.schuetzhold@uni-due.de}

\date{\today} 

\begin{abstract}
The expansion of the partition function for large coordination number $Z$ is a long standing method and
has formerly been used to describe the Ising model at finite temperatures. 
We extend this approach and study the interacting Bose gas at finite temperatures.
An analytical expression for the free energy is derived which is valid 
for weakly interacting and strongly interacting bosons.
The transition line which separates the superfluid phase from Mott insulating/normal gas phase
is shown for fillings $\langle\hat n\rangle=1$ and $\langle\hat n\rangle=2$.
For unit filling, our findings agree qualitatively with Quantum Monte-Carlo results. 
Contrary to the well-known mean-field result, the shift of the critical temperature 
in the weakly interacting regime is apparent.
%Finally, 
%we discuss the inclusion of higher order term such as the self avoiding walks 
%that alter the resummation of the ring diagram as an improvement of the convergence towards the exact transition 
%line.
\end{abstract}


\maketitle

\section{Introduction}

The Bose-Hubbard model describes interacting bosons in periodic
potentials and was originally introduced to study low energy excitations 
and the solid-superfluid phase transition in $^{4}$He \cite{GK63,GT65,FG66}.
Later on it has been used to describe properties of Josephson junction arrays \cite{GPGVM89}
and $^{4}$He absorbed in porous media \cite{FWGF89}.
Within the last decade, the Bose-Hubbard model has attracted increasing 
attention since its experimental realization became possible 
by means of interacting bosons in optical lattices \cite{G02,B10,J98,Z03,B05,S07,RSN97}.
The system exhibits a Mott insulating and a superfluid phase.
The second order phase transition, resulting from the competition 
between hopping energy and interaction energy, has been observed  
for bosons in optical lattices \cite{B10}.

On the theoretical side, the Bose Hubbard Hamiltonian has 
been studied perturbatively for strong and weak coupling \cite{FM94,FM96,DZ06,THH09,SD05,SKPR93}
and for large and small filling factors \cite{SUXF06,FSU08}.
Furthermore, it has been investigated using 
dynamical mean-field theory \cite{KZ93,HSH09,LBHH11,LBHH12,AP98}, 
a slave-boson approach \cite{DODS03}, exact diagonalisation \cite{K16}, Monte Carlo
methods \cite{TP10,CPS07,CGPB08} and the Density Matrix Renormalization Group technique \cite{CBPE12,E11,EFG11}.

The goal of the present work is to study
thermal properties of interacting Bosons
using an expansion of the partition function into inverse 
powers of the coordination number $Z$, i.e the number 
of nearest neighbors on a given lattice.
The formalism has been previously applied to the Ising model \cite{B59,B60,HC61,S73a,S73b}
where $Z$ was taken to be the number 
of spins in the range of the exchange potential.
It was found that the results are 
in good agreement with rigorous high-temperature and 
low-temperature expansions \cite{HC61}.
A related approximation method was previously applied
to investigate the dynamics of lattice site correlations
after a quantum quench \cite{NS10,QKNS14,KNQS14} and particle-hole pair creation in tilted optical lattices \cite{QNS11}.
The $1/Z$-expansion is valid for all values 
of the interaction energy and the hopping energy.
Therefore, it is suited to study the intermediate regime
between the strongly interacting regime on the one 
hand and the weakly interacting regime on the other hand.
In the strongly interacting regime, the (thermal) state of the system 
respects the $U(1)$-symmetry of the Hamiltonian and the 
correlations are short-ranged.
If the kinetic term dominates, the 
$U(1)$-symmetry is broken and spatial correlations 
become long-ranged.
It should be noted, that the $U(1)$-phase is divided into a Mott insulating phase and a normal gas phase
which are separated by a crossover.
This crossover occurs when the typical energy of thermal excitations, $k_BT$, is 
of the same order as the energy gap of the system.
It becomes manifests in the gradually vanishing of the Mott lobes \cite{DLVW05,G99,PK09}.

In the following we will focus on the finite temperature phase transition 
of the Bose-Hubbard system in three dimensions \cite{PK09,CGPB08,YC05,PKHT08,TP10}.
The paper is organized as follows. 
After reviewing the $1/Z$-expansion of the partition function for the classical Ising model 
in section \ref{cumulant}, 
we apply it to interacting Bosons in section \ref{BH} and derive various thermodynamic quantities. 
In section \ref{phase}, we depict finite temperature phase diagrams which are evaluated 
from the analytical expressions and compare with results from the literature.

\section{The cluster expansion}\label{cumulant}

We start by reviewing the main arguments of \cite{B59,B60}
which led to a systematic expansion of the Ising model 
partition sum into inverse powers of the coordination number $Z$.
%
%How the same idea can be applied to the Bose-Hubbard model,
%will be discussed in the next section.
%
The energy of the classical Ising model with a constant 
interaction strength $V$  within a finite range is given by 
\begin{align}
E=-\frac{1}{2}\frac{V}{Z}\sum_{\mu\neq\nu}T_{\mu\nu} s_\mu s_\nu\,,
\end{align}
where $s_\mu$ is a random variable which take the values $\pm1$
and $T_{\mu\nu}$ is equal to unity inside the interaction range 
and zero elsewhere.

\begin{figure}[ht]
%\begin{widetext}
\begin{center}
\includegraphics[width=8cm]{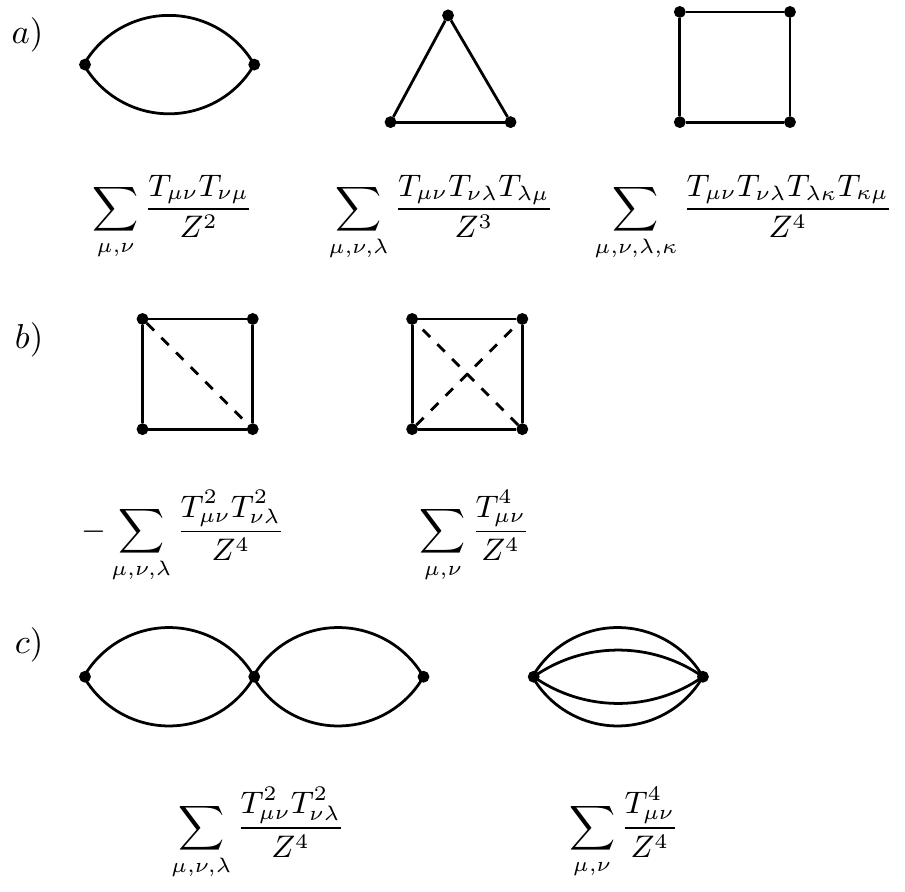}
% \includegraphics[width=17cm,bb=0 0 872 354]{ringtex.pdf}
 % ringp.pdf: 872x354 pixel, 72dpi, 30.76x12.49 cm, bb=0 0 872 354
\end{center}
%\end{widetext}
\caption{$a)$ Graphical illustration of the first three geometrical factors of order $1/Z$
in the partition sums of the Ising model (\ref{freeen}) and 
the Bose-Hubbard model (\ref{Free}) up to the 4th order. The lines represent 
the hopping between the lattice sites whereas the points 
represent the correlations at a single site.
The sums are not restricted such that the 
Fourier transform can be performed.
For this reason, it is necessary to add additional
diagrams which are at least of order $1/Z^2$.
In $b)$ we depicted the terms which have to be 
added to the four-vertex expression in $a)$.
The dashed lines illustrate the identification
of the connected vertices. 
Each dashed line in a diagram increases the order of 
the diagram by a factor $1/Z$.
$c)$ The geometrical factors of these higher order diagrams are the same 
as in $b)$ but the correlations functions are different 
since also four lines meet at a vertex.
%Diagram set up to the 4th order. The lines represent the hopping 
%between sites: 
%$T_{\mu\nu}$ while the vertices represent the $2n$- points correlations  
%$\langle T(\hat X_\mu^\dagger (\beta_1) \dots \hat X_\mu^\dagger (\beta_n) \hat 
%X_\mu (\beta'_1)
%\dots \hat X_\mu (\beta'_n))\rangle$ where $X_\mu=s_\mu$ for the spin case and  
%$\hat b_\mu$ for the Bose gas case. 
%A dashed line is drawn to connect two vertices corresponding to the same site. 
%The diagrams with dashed lines compensate the corresponding 
%one without dashed lines by subtracting the contributions of 
%two vertex corresponding to the same site. 
%A diagram containing a vertex with more than two legs 
%contribute at least with a power $1/Z$ than those involving two legs only. 
%Similarly, the presence of a dashed line in a diagram restricts the counting 
%of site by a factor $1/Z$ and accounts for self-avoiding closed paths in the 
%expansion.  The net result of this analysis is that, by adding the extra factor 
%$Z^{-n}$ to each power $n$ in the expansion, only the 
%ring 
%diagrams without a dashed line are the leading corrective terms that must be 
%resumed.
}\label{diagrams}
\end{figure}

The Curie point is reached by varying the spin density at fixed interaction
strength or by increasing the interaction strength at fixed 
density.
When the number of spins in the interaction range is increased whereas
the potential strength is decreased $\sim 1/Z $, the Curie point 
is fixed.
Assuming that only configurations with an average magnetization 
$R=\sum_\mu s_\mu/N$ are taken into account, we obtain
for $Z\rightarrow N$ the molecular field limit
\begin{align}
E=\frac{N}{2}V R^2 +\mathcal{O}(1)\,.
\end{align}
This limiting procedure motivates 
the choice of $1/Z$ as a suitable expansion parameter
for calculating corrections to the molecular field limit.
The corrections are determined by the correlations among
different spins.
To quantify the correction up to first order $1/Z$, we consider 
the partition function $\Xi$ and the free energy $\ln \Xi$, respectively.
\begin{align}\label{freeen}
\ln \Xi=\ln \sum_{\{s_\mu=\pm1\}} \exp\left[\frac{\beta V}{2Z} \sum_{\mu\neq\nu}T_{\mu\nu}s_\mu s_\nu \right]
\end{align}
This function  can be represented as  a sum of connected diagrams and 
its  
expansion (\ref{freeen}) into powers of $\beta$ produces 
\begin{align}
\ln \Xi=\sum_n \frac{\beta^n}{n!}M_n\,,
\end{align}
where the $M_n$'s can be ordered by powers of the expansion parameter $1/Z$.
For simplicity, we assume a vanishing magnetization, $R=0$, such that 
the leading contribution is of order $1/Z$.
Diagrammatically, the lowest order in $1/Z$ can be represented 
to each order in $\beta$ by a ring diagram, see Fig.~\ref{diagrams}.

%\begin{widetext}
%\end{widetext}

%
Other diagrams, for example the ladder diagram in Fig.~\ref{diagrams},
are of higher order since each solid bond gives an additional factor of $1/Z$.
The ring diagrams correspond to the summation over all 
self-avoiding closed paths on the lattice,
\begin{align}\label{ringsum}
M_n^\mathrm{ring}=\frac{(n-1)!V^n}{2Z^n}\sum_{\mu_1\neq\mu_2\neq...\neq\mu_n}
T_{\mu_1\mu_2}T_{\mu_2\mu_3}....T_{\mu_n\mu_1}\,. 
\end{align}
Since it is difficult to evaluate the sums for large $n$, 
even for nearest-neighbor interaction on a square lattice, we turn to the so-called 
random-phase approximation.
Therefore we include also closed paths which cross two or more lattice 
sites more than once such that the summations in (\ref{ringsum})
are unrestricted.
This procedure is formally justified since the additional terms, represented by 
diagram with dashed lines in Fig.~\ref{diagrams}, 
are of higher order in $1/Z$.
However, it should be noted that this appoximation leads to 
a severe violation of a sum rule $\sum_\mu s_\mu^2=N$.
Using the Fourier transform given by 
\begin{align}\label{hopping}
T_{\mu\nu}=\frac{Z}{N}\sum_\mathbf{k}T_\mathbf{k} e^{i\mathbf{k}\cdot(\mathbf{x}_\mu-\mathbf{x}_\nu)}
\end{align}
we arrive at the expression
\begin{align}
M_n^\mathrm{ring}&=\frac{(n-1)!V^n}{2Z^n}\sum_{\mu_1...\mu_n}T_{\mu_1\mu_2}....T_{\mu_n\mu_1}+\mathcal{O}(1/Z^2)\nonumber\\
&=\frac{(n-1)!V^n}{2}\sum_\mathbf{k}T_\mathbf{k}^n+\mathcal{O}(1/Z^2)\,.
\end{align}
Summing all the contributions of the ring diagrams, one
obtains the free energy
\begin{align}\label{partspin}
\frac{1}{N}\ln \Xi=-\frac{1}{2N}\sum_\mathbf{k}\ln[1-\beta VT_\mathbf{k}]+\ln 2+\mathcal{O}(1/Z^2)\,.
\end{align}
From here follows that the correlation length diverges 
if the temperature approaches the Curie temperature 
$T_C=1/V$.
A detailed discussion of (\ref{partspin}) can be found in \cite{B59,B60}.

\section{Bose-Hubbard model}\label{BH}

\begin{figure}[t]
\includegraphics{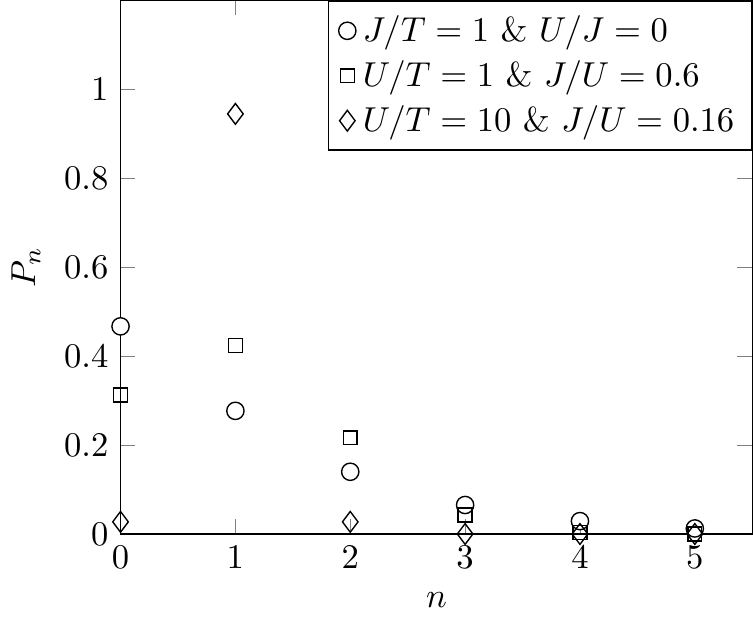} 
\caption{Probability distribution at unit filling for the 
free Bose gas (circles) and the interacting Bose gas, (squares and diamonds),
see also equations (\ref{prob}) and (\ref{probbose}).
}\label{probabilities}
\end{figure}

Inspired from the previous section, we use the same expansion technique to analyze the 
Bose-Hubbard model up to the first order in $1/Z$.
The grand canonical Bose-Hubbard-Hamiltonian has the form
\begin{align}\label{BHhamilt}
\hat{H}&=\sum_\mu \left[\frac{U}{2}\hat{n}_\mu(\hat{n}_\mu-1)-\mu 
\hat{n}_\mu\right]
-\frac{J}{Z}\sum_{\mu,\nu}T_{\mu\nu}\hat{b}^\dagger_\mu\hat{b}_\nu\nonumber\\
&=\sum_\mu H_\mu+\frac{1}{Z}\sum_{\mu\nu}\hat{H}_{\mu\nu}-\mu \hat{N}\,, 
\end{align}
where $U$ denotes the on-site repulsion, $J$ is the hopping rate, and 
$\mu$  the chemical potential.
By analogy, we shall consider $T_{\mu\nu}=1$ for next-neighbors and zero elsewhere renormalized 
with factor $1/Z$.
Switching to the interaction picture,
the free energy takes the form
\begin{align}
\ln \Xi
      =&\ln{\rm Tr}\Bigg(e^{-\beta (\sum_\mu\hat H_\mu-\mu \hat N)}\nonumber\\
      &\times {T}\left\{\exp\left[-\int_0^\beta d\beta' 
\frac{1}{Z}\sum_{\mu\nu}\hat H_{\mu\nu}(\beta')\right]\right\}\Bigg)\,,
\end{align}
The hopping term depends on imaginary time 
according to
$\hat H_{\mu\nu}(\beta)=e^{\beta(\sum_\mu\hat H_\mu-\mu \hat N)} \hat H_{\mu\nu} e^{-\beta(\hat 
\sum_\mu\hat H_\mu-\mu \hat N)}$
and
${T}(\dots)$ denotes the imaginary time ordering operator.
As in field theory, the disconnected
diagrams are absent in the logarithm 
since the free energy is the generating functional of 
all connected Green's functions \cite{PS95}.
Using the series representation of the 
exponential, we find
\begin{align}\label{freeenergy}
\ln \left(\frac{\Xi}{\Xi_0}\right)&=\left\langle  \sum_{n=1}^\infty \frac{(-1)^n}{n!Z^n} 
T\left(\int_0^\beta d\beta' \sum_{\mu\nu} \hat H_{\mu\nu}(\beta')\right)^n 
\right\rangle_\mathrm{con}
\end{align}
where the brackets in (\ref{freeenergy}) 
corresponds to the thermal expectation values at $J\rightarrow0$,
\begin{align}
\langle ...\rangle=\frac{\mathrm{Tr}\left(...e^{-\beta \left(\sum_\mu\hat H_\mu-\mu \hat N\right)}\right)}{\Xi_0}
\end{align}
with the zeroth order partition function
\begin{align}
\Xi_0&=
\left(\sum_{n=0}^\infty e^{-\beta[Un(n-1)/2-\mu n]}\right)^N\,.
\end{align}
The connected parts can be ordered by powers of $1/Z$.
In first order only the ring diagrams contribute to the partition function (see Fig.\ref{diagrams}
and compare with the corresponding expression (\ref{ringsum})),
\begin{widetext}
\begin{align}\label{Free}
\ln \left(\frac{\Xi}{\Xi_0}\right)
&=\sum_{n=1}^\infty \frac{(-1)^n}{nZ^n} 
\sum_{\mu_1\neq\mu_2 \dots\neq \mu_n}
\int_0^\beta d\beta_1 \dots  \int_0^\beta d\beta_n 
\left\langle T\left( \hat H_{\mu_1\mu_2}(\beta_1)\hat H_{\mu_2\mu_3 }(\beta_2)
\dots \hat H_{\mu_n \mu_1}(\beta_n)\right)\right\rangle + {\cal O }(1/Z^2)\,.
\end{align}
\end{widetext}
Since the summation runs only over pairwise distinct lattice sites,
the thermal expectation value can be expressed in terms of 
the on-site Green functions
\begin{align}
G(\beta_1-\beta_2)&=\langle T(\hat b^\dagger_\mu(\beta_1) \hat b_\mu (\beta_2)) \rangle\nonumber\\
&=\sum_{l=-\infty}^\infty e^{2\pi i (\beta_1-\beta_2)l/\beta}\tilde{G}(i2\pi l/\beta)\,.
\end{align}
Further calculations identify these components as:
\begin{align}\label{greensfourier}
\tilde{G}(i2\pi l/\beta)&=\int_0^\beta d\beta' G(\beta') 
 e^{-i \frac{2\pi l \beta'}{\beta}}\nonumber\\
& =\sum_{n=0}^\infty 
\frac{(n+1)(p_{n}-p_{n+1})}{2\pi i l/\beta -  Un +\mu}
\end{align}
with the $p_n$ being the on-site occupation probabilities 
in absence of Hopping,
\begin{eqnarray}\label{mf}
p_n=\frac{e^{-\beta\left(\frac{U}{2}n(n-1)-\mu n\right)}}{\sum_{m=0}^\infty 
e^{-\beta\left(\frac{U}{2}m(m-1)-\mu m\right)}}\,
\end{eqnarray}
Again, we apply the random-phase approximation and include
terms of order $1/Z^2$ such that the 
sum over all lattice sites can be performed. 
This allows to to bring the series into the tractable form
\begin{align}\label{randomphasesum}
&\ln \left(\Xi/ \Xi_0\right)\nonumber\\
&=
\sum_{n=1}^\infty
\sum_{\mu_i}\sum_{l=-\infty}^\infty
T_{\mu_1\mu_2}
\dots T_{\mu_n \mu_1} \frac{(-J)^n\tilde{G}^n\left(\frac{i2\pi l}{\beta}\right)}{nZ^n}
+\mathcal{O}\left(\frac{1}{Z^2}\right)\nonumber \\
&=-\sum_{\vc{k}}\sum_{l=-\infty}^\infty
\ln\left(1+JT_\vc{k} \tilde{G}(i2\pi l/\beta)\right)+\mathcal{O}\left(\frac{1}{Z^2}\right)\,.
\end{align}
Here, the Fourier transform of $T_{\mu\nu}$ has been defined
as in (\ref{hopping}).
The infinite sum over $l$ can be converted
to an integral along the real axis
according to
\begin{align}\label{principal}
\ln \Xi&=-\beta\sum_\mathbf{k}\mathrm{Im}\int_{-\infty}^\infty \frac{d\omega}{\pi}
\mathcal{P}\frac{\ln(1+JT_\mathbf{k}\tilde{G}(\omega+i0))}{\exp(\beta\omega)-1}\nonumber\\
&+N\ln\left(\sum_{n=0}^\infty e^{-\beta[Un(n-1)/2-\mu n]}\right)+\mathcal{O}(1/Z^2)\,,
\end{align}
where the Bose factor has simple poles at $\omega=i2 \pi l/\beta$.
The compact analytical expression (\ref{principal}) for the free energy 
of the interacting Bose gas is the main result of our paper.
Since we did not introduce a symmetry breaking parameter 
in the Hamiltonian, the partition function 
is valid above the critical temperature for a given ratio $J/U$
It is valid in the strong interacting regime, $J/U\ll1$, 
and gives also the correct limit in the non-interacting limit, $U/J\rightarrow 0$.
In particular, we deduce from (\ref{principal}) 
the partition function of the free Bose-gas,
\begin{align}
\lim_{U\rightarrow 0}\ln \Xi=-\sum_{\vc{k}}\ln\left(1-\exp[\beta(\mu+JT_\vc{k})]\right)\,.
\end{align}
The on-site probability distribution can be obtained by
replacing the potential energy in (\ref{principal})
according $Un(n-1)/2\rightarrow u_n$ and taking the 
partial derivative w.r.t.~$u_n$,
\begin{align}\label{prob}
P_n &=-\frac{1}{\beta N}\frac{\partial}{\partial u_n}\ln\Xi|_{u_n=Un(n-1)/2}
\nonumber \\
&= p_n+\frac{1}{N}\sum_\mathbf{k}\sum_{l=-\infty}^{\infty}
\frac{JT_\mathbf{k}}{1+JT_\mathbf{k}\tilde{G}(i 2\pi l/\beta)}
\bigg[p_n\tilde{G}{(i 2\pi l/\beta)} \nonumber\\
&-\frac{(n+1)p_n}{i2\pi l/\beta-U n+\mu} 
+\frac{np_n}{i2\pi l/\beta-U (n-1)+\mu}\nonumber\\
&+\frac{n(p_{n-1}-p_n)/\beta}{(i2\pi l/\beta-U(n-1)+\mu)^2}
-\frac{(n+1)(p_n-p_{n+1})/\beta}{(i2\pi l/\beta-Un+\mu)^2}\bigg]\,.
%\nonumber\\
%&=p_n\nonumber\\
%&+{\rm Im} \frac{1}{N}
%\sum_{\vc{k}}\int_{-\infty}^{\infty}\frac{d\omega}{\pi}\frac{1}{
%\exp(\beta\omega)-1}\frac{1}{(1+JT_\mathbf{k}\tilde{G}(\omega+i0))}
%\nonumber \\
%&\times\bigg[
%\frac{JT_\vc{k}n(p_{n-1}-p_{n})}{(\omega+i0 - U(n-1) +\mu)^2}-
%\frac{JT_\vc{k}(n+1)(p_{n}-p_{n+1})}{(\omega+i0 - Un +\mu)^2}
%\nonumber \\
%&-p_{n}\bigg(\beta-\frac{\beta JT_\vc{k}n}{\omega+i0 - U(n-1) +\mu}+
%\frac{\beta JT_\vc{k}(n+1)}{\omega+i0 - Un +\mu} \bigg)
%\bigg]\,.
\end{align}
From this it is possible to derive an analytical expression
for the probability distribution of the free 
Bose gas which has never been obtained before, 
\begin{align}\label{probbose}
\lim_{U\rightarrow 0}P_n&=p_n\\
&+(e^{\beta \mu}(n+1)-n)\frac{p_n}{N}
\sum_{\vc{k}}\left(\frac{1-e^{-\beta\mu}}{e^{-\beta(JT_\vc{k}+\mu)}-1}+1\right)\,.\nonumber
\end{align}
In Fig.~\ref{probabilities}, the probability distribution is depicted for 
various parameters.
%
%The nonperturbative expression (\ref{principal})
%allows some interpretation of the partition function.
%
%$\Xi$ assigns to every closed walk within an imaginary time 
%$\beta$ a certain weight.
%
%This weight depends on the length \textit{and} the 
%number of its self-intersections.
%
%At lowest order we have only self-avoiding ring diagrams.
%
%The weight of closed walks with the same length but and a finite 
%number of knots will be different since the correlations 
%feel whether the site has been visited before or not due 
%to the onsite interaction.
%
%In the limite $U=0$
%the weight does only depend on the length of the 
%path and \textit{not} on the number of self-intersections
%of a given path.
%
%This is the reason why in the limit $U/J\rightarrow0$
%the partition function becomes exact in the random phase approximation 
%which corresponds to the summation over all closed walks irrespective 
%of their number of intersections,
%
The filling can be deduced from the partition function $\Xi$ by
taking the partial derivative w.r.t.~the chemical potential,
\begin{align}\label{n}
\langle \hat{n}\rangle&=\frac{1}{\beta N}\frac{\partial}{\partial\mu}\ln\Xi=
\sum_{n=0}^\infty n P_n\nonumber\\ 
 &=\sum_{n=1}^\infty np_n-\frac{1}{N\beta}\sum_\mathbf{k}
 \sum_{l=-\infty}^{\infty}
 \frac{JT_\mathbf{k}\partial_\mu \tilde{G}(i2\pi l/\beta)}
 {1+JT_\mathbf{k}\tilde{G}(i2\pi l/\beta)}\,.
\end{align}
The two-point correlations are deduced via 
the partial derivative w.r.t.~the $T_\mathbf{k}$
\begin{align}
\langle\hat{b}_\mu^\dagger\hat{b}_\nu\rangle=
\frac{Z}{J}\frac{\partial}{\partial T_{\mu\nu}}\ln\Xi
=\frac{Z}{J} \sum_{\mathbf{k}}\frac{\partial T_{\mathbf{k}}}{\partial T_{\mu\nu}}
\frac{\partial}{\partial T_{\mathbf{k}}}\ln\Xi\,,
\end{align}
%
%In other words, the hierarchical expansion is 
%equivalent to the expansion of the partition 
%function in section (\ref{BH}) (at least to first order 
%in $1/Z$).
which leads to
\begin{eqnarray}\label{corr}
\langle\hat{b}_\mu^\dagger\hat{b}_\nu\rangle&=&
\frac{i}{2\pi}\lim_{\epsilon\rightarrow 0^+}\int_{-\infty}^\infty
\frac{d\omega}{e^{\beta \omega}-1}\frac{1}{N}\sum_{\mathbf{k}}
e^{i\mathbf{k}\cdot(\mathbf{x}_\mu-\mathbf{x}_\nu)}
\nonumber \\
&&
\left(\frac{\tilde{G}(\omega+i\epsilon)}{1+J T_\mathbf{k}\tilde{G}(\omega+i\epsilon)}-
\frac{\tilde{G}(\omega-i\epsilon)}{1+J T_\mathbf{k}\tilde{G}(\omega-i\epsilon)}\right)
\,.
\end{eqnarray}
%Note that this last expression can be alternatively deduced from the hierarchical 
%formalism \cite{NS10} which is equivalent to the random phase approximation (see appendix). 
Combining the results (\ref{n}) and (\ref{corr}), we obtain using the Fourier transform the momentum distribution,
\begin{eqnarray}\label{nk}
n_{\mathbf{k}}=
\langle \hat{n}\rangle +\sum_{\mathbf{x}_\mu\not=\mathbf{x}_\nu} 
e^{i\mathbf{k}.(\mathbf{x}_\mu-\mathbf{x}_\nu)}
\langle\hat{b}_\mu^\dagger \hat{b}_\nu\rangle \,.
\end{eqnarray}
The partition function (\ref{principal}) can also be used to 
determine the energy, the specific heat or the entropy 
of the system.

\section{Quantum phase transition}\label{phase}

The Bose gas in a lattice resides in the Mott insulating
or normal gas phase when the interaction is dominating over 
the kinetic hopping terms.
In this phase, the correlations are short-ranged and 
the energy spectrum is gapped.
When the hopping energy $J$ increases,
the correlation length grows and 
diverges at a critical value $J_\mathrm{c}$.
In the following, we use the divergence of the correlation 
for determining the phase boundary.

\begin{figure}[t]
\includegraphics{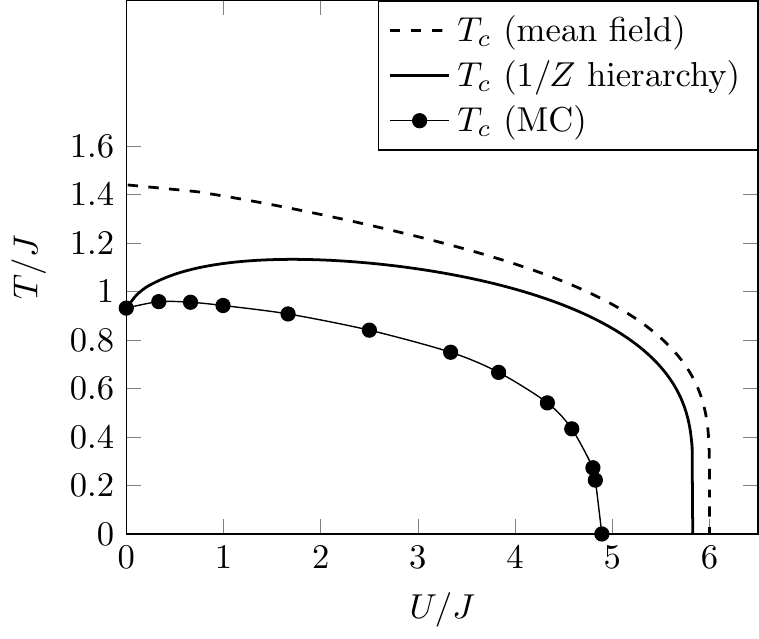} 
\caption{Phase diagram of the 3D-Bose-Hubbard model
for $\langle\hat{n} \rangle=1$ using the $1/Z$-results. We compared our results with
the mean-field result and with Monte-Carlo-calculations, reproduced from \cite{CPS07}.
The phase diagram has been compared with experimental data in \cite{SPMR15}.}\label{phasediagramN1}
\vspace{1cm}
\includegraphics{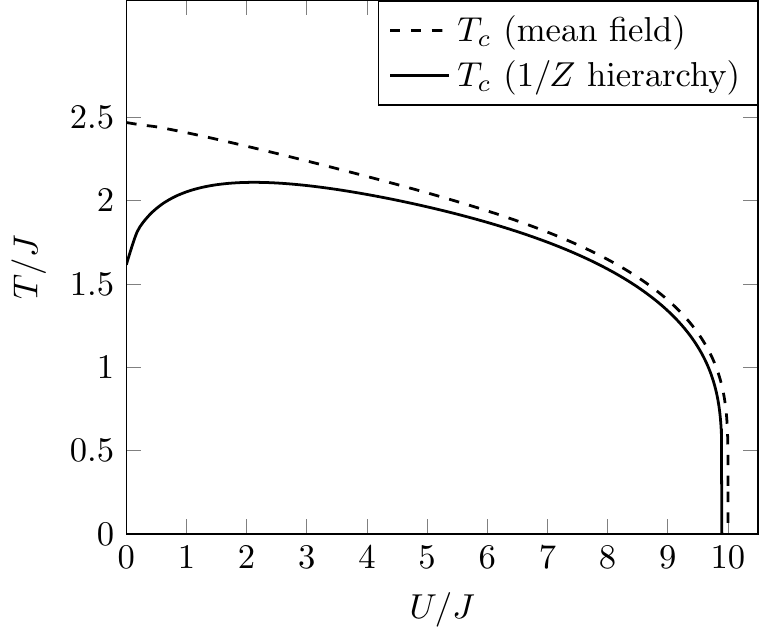} 
\caption{Phase diagram of the 3D-Bose-Hubbard model
for $\langle\hat{n} \rangle=2$.}\label{phasediagramN2}
\end{figure}

\begin{figure}
\begin{minipage}{\columnwidth}
\includegraphics{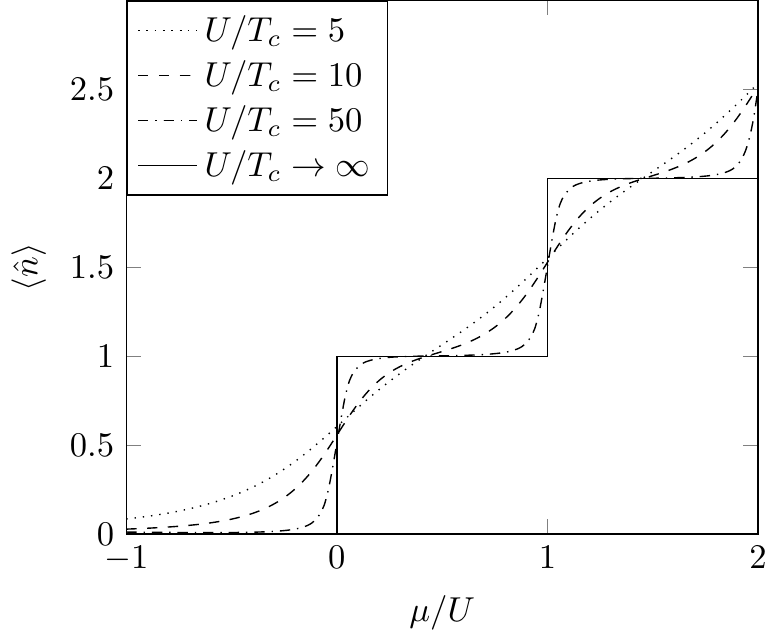} 
 \end{minipage}
\begin{minipage}{\columnwidth}
\rule[0ex]{3.3pt}{0pt}\includegraphics{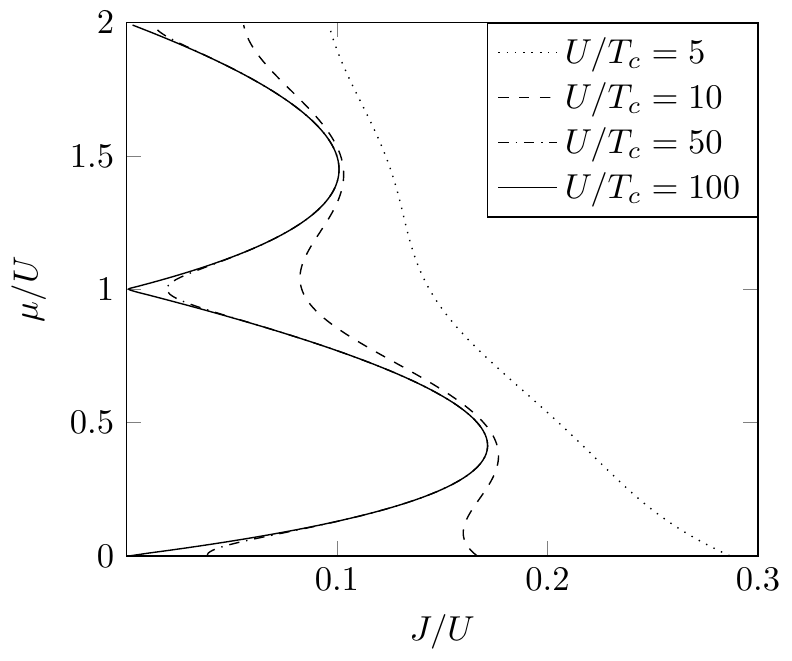}
 \end{minipage}

\caption{\textit{Top}: Expectation value of the number operator for 
different critical temperatures. At each point, the
critical hopping is determined by the relation (\ref{meanfield}). In the limit $T_c\rightarrow 0$,
the curves converge to a step-like function where only 
integer fillings are allowed. 
The curves soften when the temperature is increased and 
at $U/T_c\approx 5$ the inflexion points are vanishing.
This is the crossover between Mott insulating and normal gas phase \cite{DLVW05,G99}.
\textit{Bottom}: The solid line 
encloses the Mott lobes for filling equal to $\langle \hat{n}\rangle=1$ and 
$\langle \hat{n}\rangle=2$ in the limit of vanishing temperature. 
For increasing temperature, the Mott lobes are 
vanishing gradually.}\label{numberoperator}
\end{figure}

%At zero temperature, the fluctuations and correlations 
%of the Bose-Hubbard system are controlled by the 
%ratio $J/U$.
%

%For finite temperatures, there are also thermally 
%activated particle-hole excitations which determines 
%the transition point between normal liquid phase 
%and superfluid phase.
%
%The normal liquid phase has the same symmetry as the 
%Mott-insulating phase and differs only from it 
%by the density of the particle-hole excitations \cite{CPS07}.

%From the free energy (\ref{principal})
%we can obtain the phase-boundary at finite temperatures 
%within the random-phase approximation.
%
To first order in $1/Z$ within the random phase approximation, criterion
is equivalent to the divergence of the density distribution 
$n_{\mathbf k}$ at $\mathbf k=\mathbf 0$.
This leads to the finite temperature mean-field result \cite{BV04,KPG06,FWMGB06},
\begin{align}\label{meanfield}
1+J^\mathrm{RPA}_\mathrm{c} \tilde{G}(\omega=0)=0\,.
\end{align}
For zero temperature, we obtain from (\ref{meanfield})
\begin{align}
J^\mathrm{RPA}_\mathrm{c}=U(3-2\sqrt{2})\approx 0.171 U\,.
\end{align}
The improvement over the well-known mean-field phase 
diagrams can be achieved 
if the dependence of the filling on the parameter $J/U$
is taken into account.
Therefore, we calculate the expectation value of the particle number~(\ref{n}) for a given 
critical temperature and the corresponding 
critical hopping energy $J^\mathrm{RPA}_c$.
From the intersections of this curve with $\langle\hat{n} \rangle=1$
and $\langle\hat{n} \rangle=2$ at various temperatures, we can 
determine the phase boundary at a given filling, see 
Fig.~\ref{phasediagramN1} and Fig.~\ref{phasediagramN2}.
Since the partition function is exact in the limit $J/U\rightarrow\infty$,
we arrive at the correct critical temperature for the free Bose gas, $T_c/J\approx 0.93$.

As comparison, we deduced the mean field phase 
boundary using the particle number probability distribution (\ref{mf}).
As can be seen from Fig.~\ref{phasediagramN1}, the $1/Z$-method improves 
qualitatively and quantitatively the phase diagram.
First, the correct free Bose-gas limit is approached if the 
onsite interaction vanishes.
Second, there is not discontinuity in the limit $T\rightarrow0$
as in the mean field approach (see footnote in \cite{footnote}).
It is mainly due to the random phase approximation that the results
deviate from quasi-exact exact Monte-Carlo simulations \cite{CGPB08,CPS07,SPMR15}. 
Note that for filling equal to two, the finite temperature 
phase diagram has not been presented elsewhere. 

Fig.~\ref{numberoperator} (\textit{Top}) shows the filling factor $\langle \hat{n}\rangle$ as 
a function of $\mu/U$ which approaches a staircase function
in the limit of vanishing temperature.
When the temperature is increased, the plateaus vanish which 
displays the crossover between Mott insulating phase and normal 
gas phase \cite{DLVW05,G99,PK09}.
The corresponding Mott lobes for different critical 
temperatures are shown in Fig.~\ref{numberoperator} (\textit{Bottom}).

\newpage

\section{Conclusions}

We showed that the large coordination number expansion allows 
a full description of the Bose-Hubbard model at thermal equilibrium in the complete 
regime from weak to strong interactions. 
We obtained a closed expression for the partition function %partition sum 
of interacting Bosons which is valid for all regimes where the 
$U(1)$-symmetry of the model is preserved.
From the partition function we obtained local quantities such as the on-site
probabilities $p_n$ as well as the non-local thermal correlations 
$\langle\hat{b}_\mu^\dagger\hat{b}_\nu\rangle$. 
It should be noted that, in the limit of 
vanishing temperature, the correlation 
functions $\langle\hat{b}_\mu^\dagger\hat{b}_\nu\rangle$ 
as well as the occupation probabilities 
can also be obtained from $1/Z$-hierarchy 
that describes the dynamics of lattice site correlations \cite{NS10,QNS11}.

\begin{acknowledgments}
PN gratefully acknowledges that this work was supported by the European Union's Seventh
Framework Programme (FP7-REGPOT-2012-2013-1) under grant agreement number 316165. 
Helpful discussions with Konstantin Krutitsky are gratefully acknowledged.
\end{acknowledgments}


\begin{thebibliography}{99}
\bibitem{footnote}
The discontinuity in mean field phase-diagram for $T\rightarrow0$ 
has the following origin:
%
The chemical potential is well
defined for finite temperatures and one obtains 
from (\ref{mf}) in the limit $T\rightarrow0$ the value $\mu/U=0.5$ for unit filling
which corresponds to $U/J_c=6$.
%
In contrast, the chemical potential is not well-defined 
within the Mott lobes at $T=~0$.
%
However, it is still possible to deduce the Mott lobe boundaries
from (\ref{mf}).
%
For unit filling, the tip of the 
lobe is at $U/J_c=3+2\sqrt{2}$.
%
This unphysical discontinuity ($\triangle (U/J_c)\approx 0.171$)
is not present in the $1/Z$-approach.
\end{thebibliography}
\end{document}